\documentclass[10pt, twocolumn]{IEEEtran}

\usepackage{amsmath}
\usepackage{amssymb}    % mathbb
\usepackage{amsfonts}
\usepackage[english]{babel}
\usepackage{cite} %reference contraction
\usepackage{multirow}
\usepackage{stfloats}    % big equations float

\usepackage{algorithm}   %format of the algorithm
\usepackage{algorithmic} %format of the algorithm

\usepackage{caption} % Figure to Fig.
\usepackage{graphicx}
\usepackage{epsfig}
\usepackage{subfigure} %subfigs
\usepackage{tablefootnote}
\usepackage{float}

\usepackage{url}

\usepackage{accents}
\makeatletter
\def\widebar{\accentset{{\cc@style\underline{\mskip10mu}}}}
\def\Widebar{\accentset{{\cc@style\underline{\mskip13mu}}}}
\makeatother

\newtheorem{theorem}{Theorem}

\newtheorem{proposition}{Proposition}

\begin{document}
\captionsetup[figure]{name={Fig.},labelsep=period}  %labelfont={bf},

\title{\LARGE{Weighted-Sum Average Age of Information\\
 in Unilaterally Powered Two-Way Data Exchanging Systems}}

\author{Cheng Hu$^{\dag}$ and Yunquan~Dong$^{\dag}$  \\
$^{\dag}$Department of Electronic and Information Engineering,\\
 Nanjing University of Information Science \& Technology, Nanjing, China\\
 hucheng@nuist.edu.cn, yunquandong@nuist.edu.cn
\thanks{
This work was supported by the National Natural Science Foundation of China (NSFC) under Grant 61701247, the Jiangsu Provincial Natural Science Research Project under Grant 17KJB510035, and the Startup Foundation for Introducing Talent of NUIST under Grant 2243141701008. }
}
\maketitle

\begin{abstract}
This paper considers a two-way data exchanging system with a constant powered access point and an energy-harvesting powered smart device.
    The access point simultaneously transmits information and energy to the smart device with fixed powers $\bar{\rho} P_\text{t}$ and $\rho P_\text{t}$, respectively.
Upon collecting enough energy, the smart device performs one block of transmission immediately.
    We investigate the timeless of the two-way data exchange in terms of age of information (AoI).
Specifically, we investigate the trade-off between downlink timeliness and uplink timeliness by optimizing the weighted-sum average AoI.
    Moreover, we provide a gradient based method to compute this ratio numerically.
Our numerical results show that power-splitting based systems outperform time-splitting based systems in terms of timeliness in general.
\end{abstract}

\IEEEpeerreviewmaketitle

\section{Introduction}

In modern real-time monitoring/controlling applications such as status-sensing of environment, position/speed/ac-celeration monitoring of vehicles \cite{on-2011}, providing timely information updates is an important and critical objective of system designs \cite{queue-2014}.
    In this kind of real-time applications, however, neither traditional delay nor throughput is an adequate timeliness measure \cite{ee-2015}.
Note that when updates arrive very infrequently, the latest received update might  be generated a long time ago and hence is not fresh, even if its delay is small;
    When the throughput is high, the received updates often suffer from large queueing delays and would also be not fresh.
In \cite{rt-2012}, therefore, \textit{age of information}(AoI) which is defined as the difference between the current epoch and the generation epoch of the latest received update was proposed.

The AoI measure has been used to evaluate the timeliness of many communication systems.
%    For example, in multi-source systems \cite{taoi}, the authors formulated an AoI optimization problem and derived several general results that is applicable to a wide variety of multiple-source service systems;
%        In multi-link systems \cite{ois-2017}, the authors considered a set of links to deliver all messages and addressed the link transmission scheduling problem;
Since the AoI of an updating system is jointly determined by its arrival process and its service process, most related studies were developed based on queuing theory.
    For instance, the average AoI of $M/M/1$, $M/D/1$, $D/M/1$ queueing systems under the first-come-first-served (FCFS) policy were presented in \cite{rt-2012}.
Subsequently, the average AoI in other queuing model such as $D/G/1$ queue was given in \cite{sgo-2018}.
    Moreover, the average AoI of updating systems under other service policies such as the last-come-first-served (LCFS) policy \cite{odf-2016}, the zero-wait policy \cite{uow-2017} were also well studied.
In particular, it was shown in \cite{aoii-2018} that the LCFS policy with preemption can minimize the average AoI of updating systems.

In many practical remote monitoring systems, the battery of sensors are often limited and cannot be recharged due to their unaccessible deployments.
    To ensure the continuous and reliable functioning of these systems, wireless power transfer (WPT) have been widely used as a promising solution in \cite{wpt-2013}.
In \cite{wpu}, the authors showed that by using optimized beam-forming, the efficiency of energy transfer can be improved.
    In \cite{UAOI-2018,queue-2013}, the average AoI of a two-way data exchange system was presented, where a master node possesses the unique power supply of the system and transfers energy and data to the slave node alternatively, i.e., using the time-splitting scheme.
Although the paper has presented a full characterization of the data exchanging capability of the two-way data exchange system, the master node and the slave node cannot transmit data at the same time.

In this paper, we consider a unilaterally powered two-way data exchanging-system with \textit{power-splitting}, where an access point and a smart device exchange their own data over block Rayleigh fading channels.
    Moreover, we split the transmit power of the access point into two parts: the power used for information transmission with proportion $\bar{\rho}$ and the power for energy transfer with proportion $\rho$.
        In doing so, information transmission and energy transfer to the smart device can be performed simultaneously.
Once the smart device has collected enough energy to perform one block of transmission, it starts to transmit immediately.
    At both the access point and the smart device, we assume that new data packets are generated immediately after the transmission completion of the previous packet, i.e., following the zero-wait policy.
For both downlink and uplink transmissions, we then derive closed form average AoI.
    We also consider how the power-splitting ratio $\rho$ affects the weighted-sum average AoI of the system.
In particular, we obtained the optimal power-splitting ratio $\rho^*$ of this power-splitting system.
%%%%%%%%%%%%%%%%%%%%%%%%%%%%%%
\begin{figure*}[htbp]
  \centering
  \begin{minipage}[t]{0.49\textwidth}
    \centering
    \includegraphics[width=3.2in]{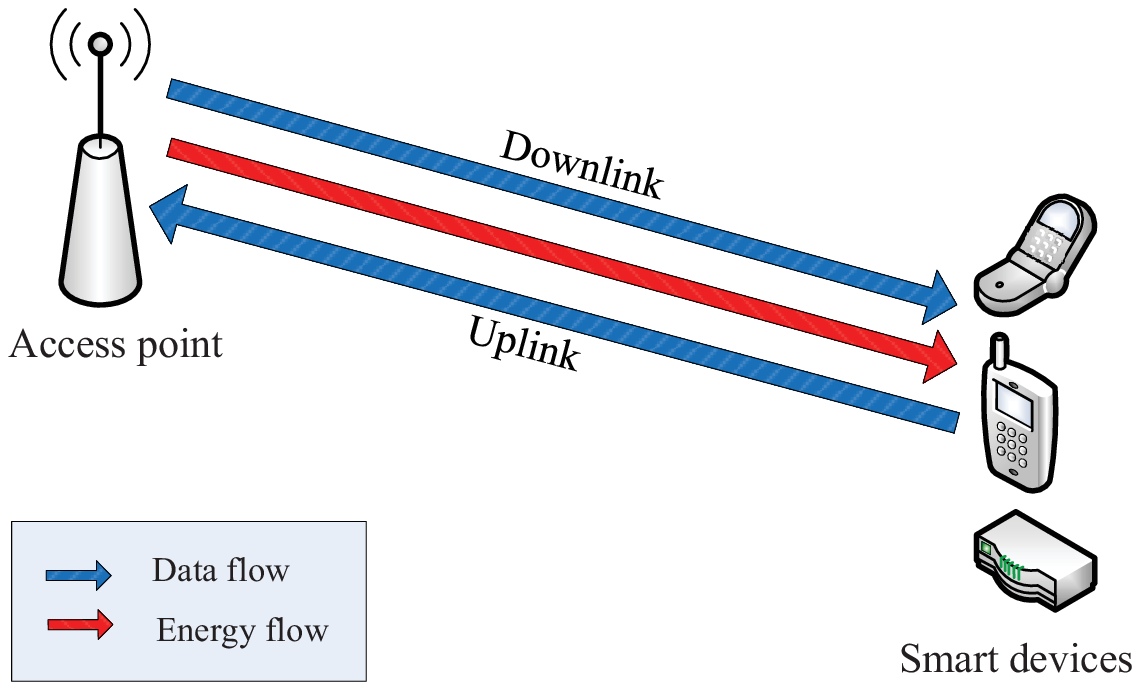}
    \caption{Two-way data exchanging system with wireless power transfer.}\label{fig:system_model}
  \end{minipage}
  \hfill
  \begin{minipage}[t]{0.49\textwidth}
    \centering
    \includegraphics[width=3.2in]{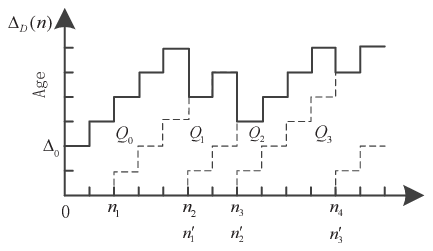}
    \caption{Sample path of downlink AoI $\Delta_D(n)$ (the upper envelop in hold).}\label{sample_path}
  \end{minipage}
\end{figure*}
%%%%%%%%%%%%%%%%%%%%%%%%%%%%%%%

The rest of the paper is organized as follows.
    In Section II, we describe the specific model of the system and the formulation of the optimization problem.
In section III, we obtain the explicit form of average AoI for both downlink and uplink communications.
    The weighted-sum average AoI is optimized over the power-splitting ratio in Section IV.
Finally, we present our simulation results in Section V and conclude the paper in Section VI.

\section{System Model}\label{sec:model}
As shown in Fig. \ref{fig:system_model}, we consider a two-way data exchange system based on power-splitting, where an access point and a smart device exchange their own data in packets via block fading channels.
    The access point has a constant power supply with power $P_\text{t}$ while the smart device does not.
The transmit power of the access point is split into two parts with a \textit{power-splitting ratio} $\rho$, where the first part at power $P_\text{d}=\bar{\rho}P_\text{t}$ is used for the \textit{downlink} information transmission and the other part at power $P_\text{e}=\rho P_\text{t}$ is used for the energy transfer to the smart device ($\bar{\rho}+\rho =1$).

\subsection{Channel and Energy Harvesting Model}

    We assume that the channels suffer from block Rayleigh-fading and additive white white Gaussian noise(AWGN), so the power gain $\gamma_n$ follows exponential distribution $f_\gamma(x) = \lambda e^{-\lambda x}$.
        We assume that the received signal-to-noise ratios (SNRs) are much smaller than unity at both the access point and the smart device.
In addition, the downlink and uplink transmissions are carried out at different frequencies bands.
    Let $T_\text{B}$ be the block length, $d$ be the distance between the access point and the smart device, $\alpha$ be the path-loss exponent, $W$ be the limited system bandwidth, and $N_0$ be the noise spectrum density.
We set the uplink transmit power the same as the average received power from downlink power transfer, i.e.,  $P_\text{u}=\frac{\rho P_\text{t}}{\lambda d^{\alpha}}$.
    Thus, the data (in nats) that can be transmitted in a block over the downlink channel and the uplink channel would be, respectively,
\begin{align}
    b_n^\text{d} &= T_{\text{B}} W \ln{ \Big( 1+\frac{\bar{\rho} P_\text{t} \gamma_n}{d^\alpha W N_0} \Big)} \approx \frac{\bar{\rho} P_\text{t} T_{\text{B}} \gamma_n}{d^\alpha N_0},   \\
    b_i^\text{u} &= T_{\text{B}} W \ln{ \Big(1+\frac{\rho P_\text{t} \gamma_i}{\lambda d^{2 \alpha} W N_0} \Big)} \approx \frac{\rho P_\text{t} T_{\text{B}} \gamma_i}{\lambda d^{2 \alpha} N_0}.
\end{align}
Hence, both $b_n^\text{d}$ and $b_i^\text{u}$ follow the exponential distribution.

We denote the energy transfer efficiency as  $\eta$, then the energy collected by the smart device in block $n$ is
\begin{equation}
    E_n = \eta \rho P_\text{t} T_{\text{B}} d^{-\alpha} \gamma_n,
\end{equation}
where the block $n$ is the period between epoch $n$ and epoch $n+1$.
    The smart device will not start transmitting data to the access point until it has collected sufficient energy for a block of transmission.
Moreover, the access point and the smart device will generate a new packet immediately after the previous packet is completely transmitted.

By definition, the downlink AoI is the difference between current epoch $n$ and the generation epoch $U_\text{D}(n)$ of the latest packet received by the smart device.
    That is
\begin{equation}
    \Delta_\text{D}(n) = n-U_\text{D}(n).
\end{equation}

Fig. \ref{sample_path} depicts a sample path of downlink AoI $\Delta_\text{D}(n)$.
    The access point generates a packet at epoch $n_1$, and its transmission is completed at  epoch $n_1'$.
When the access point detects the completion of the first packet at $n_1'$, a new packet would be generated immediately.
    It is clear that $n_1'$ and $n_2$ coincide with each other.
Thus, all the packets do not need to wait for its transmission and the waiting time is zero.
    Therefore, the  time that a packet spends in the system is equal to its service time $S_k$.

In a period of time with $N$ blocks, we suppose that the smart device receives $K$ packets.
    The data rate of the downlink transmission can then be expressed as $p=\frac{K}{N}$.
Furthermore, the average downlink AoI during this period is given by $\bar{\Delta}_{\text{D}} = \frac{1}{N} \sum_{n=1}^{N}\Delta_\text{D}(n)$.

As $N$ approaches infinity, the average downlink AoI and the downlink rate of the system are, respectively, given by
\begin{align}
    \bar{\Delta}_\text{D} &= \lim_{N \to \infty} \frac{1}{N} \left( Q_0 + \sum_{k=1}^{K-1}Q_k + \frac{1}{2} S_{K}(S_{K}+1) \right),  \label{delta_d}   \\
    p&=\lim_{N \to \infty} \frac{K}{N},  \label{p}
\end{align}
where $Q_{k}$ is the difference between the areas of the two triangles whose side length difference is $S_k$, as show in Fig. \ref{sample_path}.

Likewise, we can get the similar average uplink AoI and uplink data rate.

\subsection{Problem Formulation}

As far as the two-way data exchange system is concerned, we cannot analyze the timeliness of uplink transmission and downlink transmission in isolation.
Instead, we need to investigate the overall timeliness.
    According to the priority of the access point and the smart device, we use $w(0\leq w\leq1)$ and $\bar w$ as the weighting coefficients for the uplink transmission process and the downlink transmission process, respectively.
Since the transmission power of the access point is fixed, the information transmission and the energy transfer in the downlink are mutually restricted.
    The energy transfer process directly affects the average uplink AoI, so the average downlink/uplink AoI are also mutually constrained.
Therefore, we are interested in the following optimization problems.

\textit{Problem 1:} Minimizing the weighted-sum average AoI.
\begin{align} \label{prob:aoi}
    \mathop{\min}_{\rho}\quad &\bar{\Delta} = \bar{w}\bar{\Delta}_\text{D} + w\bar{\Delta}_\text{U},  \\
    \text{s.t.}\quad  &0 \leq \rho \leq 1, \nonumber  \\
           &0 \leq w \leq 1.  \nonumber
\end{align}

\section{Average Downlink/Uplink AoI}\label{sec:age of information}
In this section, we first investigate the statistic property of downlink service time.
    After that, we shall derive the average downlink AoI in closed form.
Finally, we study the energy harvesting process and derive the average uplink AoI in closed form.

\subsection{Downlink Service Time and Average Downlink AoI}
The \textit{downlink service time} $S_\text{D}$ is the number of blocks required to complete the transmission of a packet.
    In particular, the probability distribution $p_j^{\text{S}}=\Pr\{S_{\text{D}}=j\}$ of $S_\textrm{D}$ is given by the following proposition.

\begin{proposition}
 The probability distribution of downlink service time $S_{\text{D}}$ is given by
\begin{equation}
    p_j^{\text{S}}=\frac{(\frac{\theta}{\bar{\rho}})^{j-1}}{(j-1)!} e^{-\frac{\theta}{\bar{\rho}}},~\text{for}~j=1,2,\ldots,
\end{equation}
where
\begin{equation}
    \theta = \frac{\lambda l N_0 d^{\alpha}}{P_\text{t} T_\text{B}}.
\end{equation}
\end{proposition}

\textit{Proof:} See Appendix A.

The probability generating function (PGF) and the first two order moments of $S_\text{D}$ are given by, respectively,
\begin{align}
    G_{\text{S}}(z) &= \mathbb{E}(z^S) = ze^{\frac{\theta}{\bar{\rho}} (z-1)},  \\
    \mathbb{E}(S_{\text{D}}) &= \lim_{z \to 1^-} G_{\text{S}}'(z) = 1+\frac{\theta}{\bar{\rho}},  \\
    \mathbb{E}(S^2_{\text{D}}) &= \lim_{z \to 1^-} G_{\text{S}}''(z) + G_{\text{S}}'(z) = \Big(\frac{\theta}{\bar{\rho}}\Big)^2+\frac{3\theta}{\bar{\rho}}+1.
\end{align}

    Note that $\mathbb{E}(S) = 1+\frac{\theta}{\bar{\rho}}$ is the average downlink service time of  packets and
    $\frac{\theta}{\bar{\rho}}$ is the ratio between packet length $l$ and the average amount of information that can be transmitted in a block.

Based on the system model and the previous analysis, the average downlink AoI can be readily obtained, as shown in the following theorem.

\begin{theorem}
 If $0<\rho<1$, the average downlink AoI is, given by
\begin{align}
    \bar{\Delta}_\text{D} = 1&+\frac{\theta}{\bar{\rho}} + \frac{(\frac{\theta}{\bar{\rho}})^2+4\frac{\theta}{\bar{\rho}}+2}{2(1+\frac{\theta}{\bar{\rho}})}, \label{AoI_D}
\end{align}
where $\theta = \frac{\lambda l N_0 d^{\alpha}}{P_\text{t} T_\text{B}}$. Otherwise, it would be infinitely large.
\end{theorem}

\textit{Proof:} See Appendix B.

\subsection{Energy Harvesting Process and Average Uplink AoI}

We denote the energy collected in the $i$-th block and a period consisting of $j$ blocks as $E_i$ and $e_j$, respectively.
We then have
\begin{equation}
e_j=\sum_{i=1}^j E_i=\frac{\eta \rho P_\text{t} T_{\text{B}}}{d^\alpha} \sum_{i=1}^j \gamma_i.
\end{equation}
    According to \cite{queue-2013}, we have $f_{e_j}(x) =\frac{\mu x^{j-1}}{(j-1)!} e^{-\mu x}.$

Let $\tau_{\text{H}}$ be the number of blocks for the smart device to accumulate enough energy to perform a block of transmission.
    Since $\tau_{\text{H}}$ may occasionally be smaller than unity, we further denote   $s=\max\{1,\tau_{\text{H}}\}$.
    According to \cite{queue-2013},
\begin{align}
   \text{Pr}\{\tau_{\text{H}}=j \} &= \frac{ (\frac{1}{\eta})^j }{j!} e^{- \frac{1}{\eta}},  \\
   \label{E_s}
   \mathbb{E}(s) &=\frac{1}{\eta}+e^{-\frac{1}{\eta}}.
\end{align}

We denote the number of transmissions to complete a packet over uplink as $S$.
    Since the uplink service time $S_\text{U}$ is the sum of $S$ and the time $s_i$ to accumulate energy for each block-of-transmission, we have $S_\text{U}=\sum_{i=1}^Ss_i.$
%\label{S_U}

The PGF and the first two order moments of $S_\text{U}$ can be readily obtained as
\begin{align}
    p_j^{\text{S}} &= \frac{1}{(j-1)!} \Big(\frac{\lambda \theta d^{\alpha}}{\rho}\Big)^{j-1} e^{-\frac{\lambda \theta d^{\alpha}}{\rho}},\\
    G_\text{S}(z) &= z e^{ \frac{\lambda \theta d^{\alpha}}{\rho} (z-1) },\\
    \label{E_S}
    \mathbb{E}(S) &=1+\frac{\lambda \theta d^{\alpha}}{\rho},\\
    \label{E_S^2}
    \mathbb{E}(S^2) &=\Big(\frac{\lambda \theta d^{\alpha}}{\rho}\Big)^2 +\frac{3 \lambda \theta d^{\alpha}}{\rho}+1.
\end{align}

The moments of $s_i$ and $S_{\text{U}}$ can then be given by the following proposition.

\begin{proposition}
The first two order moments of uplink service time $S_{\text{U}}$ and $s_i$ are, respectively, given by
\begin{align}
   \label{E_s_i}
   \mathbb{E}(s_i) =&\frac{1}{\eta}+e^{-\frac{1}{\eta}},\\
   \label{E_s_i^2}
   \mathbb{E}(s_i^2) =&\frac{1}{\eta^2}+\frac{1}{\eta}+e^{-\frac{1}{\eta}},\\
   \label{E_S_U}
   \mathbb{E}(S_{\text{U}}) =&\Big(1+\frac{\lambda \theta d^{\alpha}}{\rho}\Big)\Big(\frac{1}{\eta}+e^{-\frac{1}{\eta}}\Big),\\
   \label{E_S_U^2}
   \mathbb{E}(S_{\text{U}}^2) =&\Big(1+\frac{\lambda \theta d^{\alpha}}{\rho}\Big)\Big(\frac{1}{\eta^2}+\frac{1}{\eta}+e^{-\frac{1}{\eta}}\Big)  \nonumber \\
                                 & +\left(\Big(\frac{\lambda \theta d^{\alpha}}{\rho}\Big)^2 +\frac{2 \lambda \theta d^{\alpha}}{\rho}\right) \Big(\frac{1}{\eta}+e^{-\frac{1}{\eta}}\Big)^2.
\end{align}
\end{proposition}

\textit{Proof:} See Appendix C.

%%%%%%%%%%%%%%%%%%%%%%%%%%%%%%%
\begin{figure*}[hbp]   %cross column: add *

\hspace{-6 mm}
    \begin{tabular}{cc}
    \subfigure[Average AoI versus power-splitting ratio $\rho$]
    {
    \begin{minipage}[t]{0.5\textwidth}
    \centering
    {\includegraphics[width = 3.5in] {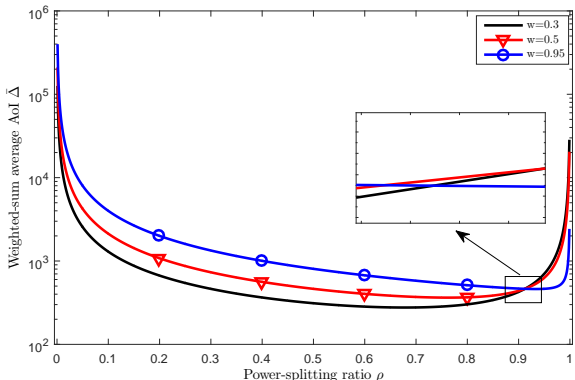} \label{fig:aoi_rho}}
    \end{minipage}
    }

    \subfigure[Optimal average AoI versus weight $w$]
    {
    \begin{minipage}[t]{0.5\textwidth}
    \centering
    {\includegraphics[width = 3.5in] {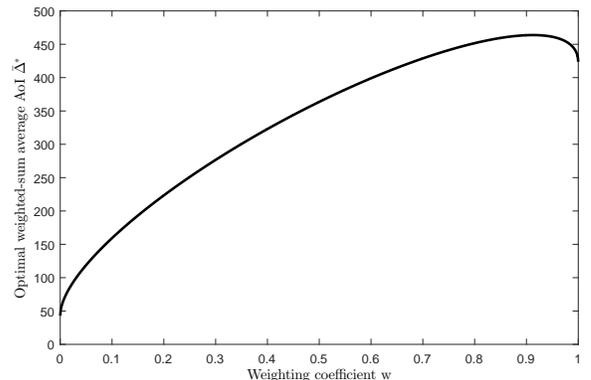} \label{fig:opt_aoi_w} }
    %{f_3_AoI_rho.eps} \label{fig:opt_aoi_rho}
    \end{minipage}
    }
    \end{tabular}
\caption{Average AoI of the power-splitting system. } \label{fig:w_aoi}
\end{figure*}
%%%%%%%%%%%%%%%%%%%%%%%%%%%%%%

Based on previous results, we get the following theorem.

\begin{theorem}
 Given that $0<\rho<1$, the average uplink AoI would be
\begin{align} \label{AoI_U}
    \bar{\Delta}_\text{U}=&\frac{3}{2} \Big(1+\frac{\lambda \theta d^{\alpha}}{\rho}\Big) \Big( \frac{1}{\eta}+e^{-\frac{1}{\eta}} \Big) +\frac{1}{2}+\frac{1}{2} \frac{1}{\eta+\eta^2 e^{-\frac{1}{\eta}}} \nonumber  \\
    &-\frac{1}{2} \frac{\rho}{\rho+\lambda \theta d^{\alpha}} \Big(\frac{1}{\eta}+e^{-\frac{1}{\eta}}\Big).
\end{align}
\end{theorem}

\textit{Proof:} See Appendix D.

\section{Downlink-Uplink Trade-off of System Freshness}\label{sec:optimization}

For a given $w$, by taking the derivative of the objective function in \textit{Problem} 1 with respect to $\rho$, we have
\begin{align}
    \frac{\partial\bar{\Delta}}{\partial\rho}=& (1-w) \left( \frac{3\theta}{2(1-\rho)^2}+\frac{\theta}{2(1+\theta-\rho)^2} \right)  \nonumber  \\
      &+w \left( -\frac{3a}{2}\frac{\lambda\theta d^{\alpha}}{\rho^2} - \frac{a}{2}\frac{\lambda\theta d^{\alpha}}{(\rho+\lambda\theta d^{\alpha})^2} \right), \label{dr:drv_aoi_rho}
\end{align}
where $a=\frac{1}{\eta}+e^{-\frac{1}{\eta}}$.

We note that for each $\rho\in[0,1]$, \eqref{dr:drv_aoi_rho} is a continuous and derivable function.
    Also, it can be readily shown that
\begin{align}
     \frac{\partial\bar{\Delta}}{\partial\rho}\Big|_{\rho=0}<0,
     \frac{\partial\bar{\Delta}}{\partial\rho}\Big|_{\rho=1}>0,
     \frac{\partial^2\bar{\Delta}}{\partial\rho^2}\Big|_{0<\rho<1}>0.
\end{align}
Therefore, there would be exactly one $\rho^\diamond\in(0,1)$ satisfying
\begin{equation}\label{rt:delta_rho_00}
\frac{\partial\bar{\Delta}}{\partial\rho}\Big|_{\rho=\rho^\diamond}=0,
\end{equation}
 which would minimize $\bar{\Delta}$.
    Since $\rho^\diamond$ would be different if the weighting coefficient $w$ is changed, we rewrite the solution $\rho^\diamond$ as a function of $w$, i.e., $\rho^\diamond(w)$.

However, explicit $\rho^\diamond(w)$ is not available in general.
    To solve \textit{Problem} 1, therefore, we have proposed an iterative algorithm based on Newton's method, as shown in \textit{Algorithm} \ref{alg:ruo_w}.
Then we have
\begin{align}\label{rt:aoi_rho_111}
    \bar{\Delta}^*=\bar{\Delta}\,|_{\rho=\rho^\diamond(w)},    &\hspace{5mm} 0 \leq w \leq 1.
\end{align}

\begin{algorithm}[!t]
\algsetup{linenosize=\small}
\scriptsize
\caption{Iterative solution to $\rho^*(w)$}
\begin{algorithmic}[1] \label{alg:ruo_w}
\REQUIRE ~~\\%Initialization
   %\STATE Function [$\bar{\Delta}$,d$\bar{\Delta}$,d2$\bar{\Delta}$] =  func1($\rho$)
    \STATE Set the initial power-splitting ratio $\rho_0$,
    \STATE Set the maximum error $\varepsilon_{\text{max}}$, the maximum number of iterations $n_{\text{max}}$;
\ENSURE ~~\\%Iteration

\WHILE{$n \leq n_{\text{max}}$, $\left|\rho(n+1)-\rho(n)\right| \leq \varepsilon_{\text{max}}$}
    \STATE update $\bar{\Delta}$ using \eqref{prob:aoi}, \eqref{AoI_D} and \eqref{AoI_U};
    \STATE $\rho(n+1) = \rho(n)-d\bar{\Delta}/d^2\bar{\Delta}$;
    \STATE $n = n+1$;
\ENDWHILE

\end{algorithmic}

\end{algorithm}

\section{Numerical Result}\label{sec:simulation}
In this section, we present the obtained results through numerical results.
    We set the system bandwidth to $W=1\text{MHz}$, the noise spectrum density to $N_0=4\times10^{-7}$, the distance between the access point and the smart device to $d=1.5\text{m}$, and the path-loss exponent to $\alpha=2$.
The total power at the access point set to $P_\text{t}=0.01$W, including the power $\bar{\rho}P_\text{t}$ for information transmission and the power $\rho P_\text{t}$ for energy transfer.
    The transmit power of the smart device is set as $P_\text{u}=\frac{\rho P_\text{t}}{\lambda d^{\alpha}}$.
The length of each packet is $l=100$ nats and the block length is $T_{\text{B}}=10^{-3}$ s.
    The Rayleigh channel parameter is $\lambda=3$.
The energy transfer efficiency is $\eta=0.5$.

%%%%%%%%%%%%%%%%%%%%%%%%%%%%%%%

\begin{figure*}[htp]   %cross column: add *

\hspace{-6 mm}
    \begin{tabular}{cc}
    \subfigure[Weighted-sum average data rate $R$ versus $p$.]
    {
    \begin{minipage}[t]{0.5\textwidth}
    \centering
    {\includegraphics[width = 3.5in] {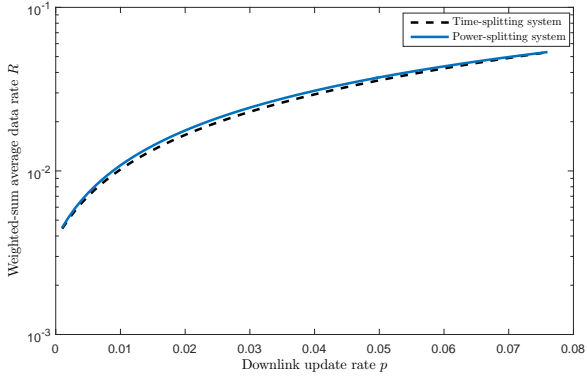} \label{fig:R_p}}
    \end{minipage}
    }

    \subfigure[Weighted-sum average AoI $\bar{\Delta}$ versus $p$.]
    {
    \begin{minipage}[t]{0.5\textwidth}
    \centering
    {\includegraphics[width = 3.5in] {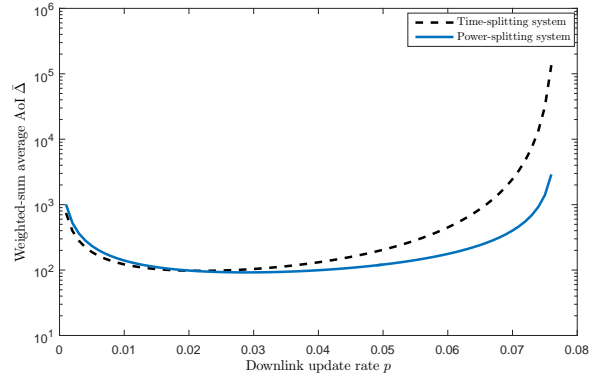} \label{fig:AoI_p}}
    \end{minipage}
    }

    \end{tabular}
\caption{Performance comparison between time-splitting systems and the power-splitting systems.} \label{fig:AoI_R_p}
\end{figure*}
%%%%%%%%%%%%%%%%%%%%%%%%%%%%%%

\subsection{Weighted-Sum Average AoI}
We present the weighted-sum average AoI $\bar{\Delta} = \bar{w}\bar{\Delta}_\text{D} + w\bar{\Delta}_\text{U}$ in Fig. \ref{fig:w_aoi}.
    %The average AoI is a convex function with power-splitting ratio $\rho$ as the independent variable.
As shown in Fig. \ref{fig:aoi_rho},  $\bar\Delta$ goes to infinity either when the power-splitting ratio $\rho$ approaches to zero or unity and achieves its minimum when $\rho$ is neither too large nor too small.
   In fact, as $\rho$ approaches zero, the power allocated to downlink information transmission would be very small, leading to large downlink service times and large downlink average AoI.
On the other hand, when $\rho$ approaches unity, the power allocated to downlink energy transfer (then for the uplink transmissions) would be  very small, leading to large period of energy collecting and large average uplink AoI.
    Moreover, we see in the left part of Fig. \ref{fig:aoi_rho} that, the larger weighting coefficient $w$ is, the more the average uplink AoI (which is large) influences $\bar\Delta$, so we have a larger $\bar\Delta$.
On right part of Fig. \ref{fig:aoi_rho}, the average uplink AoI becomes smaller since $\rho$ gets larger (i.e., more energy can be harvested).
    Thus, $\bar\Delta$ would be small when $w$ is large.
Note that, the three curves do not intersect with each other at the same point.

In Fig. \ref{fig:opt_aoi_w}, we search the minimum weighted-sum average AoI $\bar\Delta^*$ for each $w$.
    It is seen that $\bar\Delta^*$ is increasing with $w$ in the left part and is decreasing as $w$ approaches unity.
Due to the inefficiency and the randomness of the energy harvesting process, the average uplink AoI is much lager than the average downlink AoI.
    To be specific, $\bar\Delta^*$ achieve its minimum as the uplink weight $w$ is zero.
When $w$ is increased, the uplink AoI contributes more to $\bar\Delta$, which lead to a larger $\bar\Delta$.
    As $w$ approaches unity, however, $\bar\Delta$ will decrease with $w$.
This is because $\bar\Delta$ is optimized over $\rho$, which will be increased with $w$ in this case.
    Thus, the uplink AoI will be decreased.
Since the uplink AoI would dominant $\bar\Delta$ as $w$ approaches unity,  $\bar\Delta$ would also be decreased.

\subsection{Connection with Time-Splitting Two-Way Data Exchange}

It has been shown in \cite{UAOI-2018,queue-2013} that the two-way data exchanging can also be performed through time-splitting.
    To be specific, the access point generates a new packet with probability $p$ in each block.
Whenever it has data in its queue, the access point transmits its data to the smart device using all of its power.
During periods when its queue is empty, the access point transfers energy to the smart device at its full power.
    Upon receiving enough energy, the smart device can then perform information transmission to the access point.
Let $S_i$ be a period transmitting information to the smart device and $I_i$ be a period transferring energy to the smart device, the ratio between the power for energy transfer and the total power is
\begin{equation}
    \rho_\text{TS} = \lim_{N\to \infty} \frac{\sum_{i=1}^{K_\text{I}} I_i}{\sum_{i=1}^{K_\text{I}} I_i+\sum_{i=1}^{K_\text{S}}S_i},
\end{equation}
where $K_\text{I}$ is the number of periods transferring energy and $K_\text{S}$ is the number of periods transmitting data to the smart device.

Since ${\sum_{i=1}^{K_\text{I}} I_i+\sum_{i=1}^{K_\text{S}}S_i}=N$ and
\begin{align}
  \lim_{N\to \infty}\frac{1}{N}\sum_{i=1}^{K_\text{S}} S_i &= \lim_{N\to \infty}\frac{K_\text{S}}{N}\frac{1}{K_\text{S}}\sum_{i=1}^{K_\text{S}} S_i  \nonumber  \\
                                                &= p \mathbb{E}[S_i]   \nonumber  \\
                                                &= p(1+\theta),
\end{align}
where $\theta = \frac{\lambda l N_0 d^{\alpha}}{P_\text{t} T_\text{B}}$,  we have

\begin{equation}
    \rho_\text{TS} = 1-p(1+\theta).
\end{equation}

In Fig. \ref{fig:AoI_R_p}, we present the weighted-sum average AoI and data rate when the same portion of power is transferred to the smart device, i.e., $\rho=\rho_\text{TS}$.
    From Fig. \ref{fig:R_p}, we observe that the weighted-sum data rates are almost the same for the time-splitting and power-splitting system.
That is, with the same allocation of the available power between information transmission and energy transfer, we achieve the same data rate in these two systems.
    In Fig. \ref{fig:AoI_p}, we see that the weighted-sum AoI of the time-splitting system is close to that of a power-splitting system  when $p$ is small and is larger than that of a power-splitting system when $p$ is large.
Note that when $p$ is large, the weighted-sum AoI is dominated by the average uplink AoI.
    In the time-splitting system, since the uplink service time of a packet consists of the downlink busy periods (no energy transferred) and the period of energy harvesting, the corresponding variation would be larger than that of a power splitting system, which leads to larger uplink AoI.

\section{Conclusion}\label{sec:conclusion}

In this paper, we considered the timeliness and efficiency of the two-way data exchanging system with an access point and a powerless smart device.
    Using power splitting, the access point transfers a part of its energy to the smart device so that the device can transmit its own data back.
We derived closed form expressions for the average downlink and uplink AoI.
    We also investigated the weighted-sum average AoI minimizing power-splitting ratios and weighting coefficients of the system.
Our results presents a full characterization for the timeliness of the two-way data exchanging system and shed lights on the system design for two-way data exchanging with different priorities in the two directions.
    In the future, we will investigate the impact of the relay node and the digital network coding on the timeliness of the system.

\appendix

\subsection{Proof of Proposition 1}
\begin{proof}
Denote the PGF of the amount of information $b_n^\text{d}$ transmitted in a block as $f_1(x)$, and denote the \textit{p.d.f.} of the amount of information transmitted in $k$ $(k=1,2,3\ldots)$ blocks as $f_k(x)$.
    According to \cite{UAOI-2018,queue-2013}, we have following results
\begin{align}
  f_1(x) &= \frac{1}{v} e^{-\frac{x}{v}},  \\
  f_k(x) &= \frac{1}{\Gamma(k) v^k} x^{k-1} e^{-\frac{x}{v}},
\end{align}
where $v=\frac{\bar{\rho}P_\text{t} T_\text{B}}{\lambda N_0 d^{\alpha}}$.
    Thus, the probability that the downlink service time $S$ equal to $j$ which is the number of blocks is given by
\begin{align}
    p_j^{\text{S}}&=\text{Pr}\left\{\sum_{i=1}^{j-1} b_i < l,\sum_{i=1}^{j} b_i > l \right\} \nonumber \\
                     &=\int_{0}^{l}f_{j-1}(y)\mathrm{d}y \int_{l-y}^{\infty}f_1(x)\mathrm{d}x  \nonumber \\
    &=\int_{0}^{l} \frac{1}{(j-2)!v^{(j-1)}} y^{(j-2)} e^{-\frac{y}{v}} \mathrm{d}y \int_{l-y}^{\infty} \frac{1}{v}e^{-\frac{x}{v}}\mathrm{d}x  \nonumber \\
    &=\frac{(\frac{\lambda l N_0 d^{\alpha}}{\bar{\rho}P_\text{t} T_\text{B}})^{j-1}}{(j-1)!} e^{-\frac{\lambda l N_0 d^{\alpha}}{\bar{\rho}P_\text{t} T_\text{B}}}
    =\frac{(\frac{\theta}{\bar{\rho}})^{j-1}}{(j-1)!} e^{-\frac{\theta}{\bar{\rho}}},
\end{align}
where $\theta = \frac{\lambda l N_0 d^{\alpha}}{P_\text{t} T_\text{B}}.$
This proves \textit{Proposition} 1.
\end{proof}

\subsection{Proof of Theorem 1}
\begin{proof}
According to \eqref{delta_d}, \eqref{p} and Fig. \ref{sample_path}, we rewritten \eqref{delta_d} as
 \begin{align}
   \bar{\Delta}_D =& \lim_{N \to \infty} \frac{1}{N}  \sum_{k=1}^{K-1}Q_k \nonumber\\
                  =& \lim_{N \to \infty} \frac{K-1}{N} \frac{1}{K-1} \sum_{k=1}^{K-1}Q_k \nonumber \\
                  =&  p\mathbb{E}(Q_k),
 \end{align}
where $p$ is the downlink average data rate.

The average area of each $Q_k$ becomes
\begin{align}
  \mathbb{E}(Q_k) =& \mathbb{E} \left[ \frac{1}{2}(S_{k-1}+S_k)(S_{k-1}+S_k+1) - \frac{1}{2}S_k(S_k+1)  \right] \nonumber\\
                  =& \mathbb{E}^2(S_k) + \frac{1}{2}\mathbb{E}(S_k^2) + \frac{1}{2}\mathbb{E}(S_k).
\end{align}
Combining the above results, we then have
\begin{align}
  \bar{\Delta}_D &= \mathbb{E}(S_k) + \frac{1}{2} +\frac{1}{2} \frac{\mathbb{E}(S_k^2)}{\mathbb{E}(S_k)}  \nonumber \\
                 &= 1+\frac{\theta}{\bar{\rho}} + \frac{(\frac{\theta}{\bar{\rho}})^2+4\frac{\theta}{\bar{\rho}}+2}{2(1+\frac{\theta}{\bar{\rho}})},
\end{align}
which proves \textit{Theorem} 1.
\end{proof}

\subsection{Proof of Proposition 2}
\begin{proof}
In the uplink data transmission, we set the transmit power $P_{\text{U}}$ as $\frac{\rho \text{P}_t}{\lambda d^{\alpha}}$.
    Then the first two order moments of the numbers of blocks for the smart device to perform a block of transmission are, respectively, given by
\begin{align}
  \mathbb{E}(s_i) &= \sum_{j=1}^{\infty} \text{Pr}\{s=j\}j
                  = \frac{1}{\eta}+e^{-\frac{1}{\eta}},       \\
  \mathbb{E}(s_i^2) &= \sum_{j=1}^{\infty} \text{Pr}\{s=j\}j^2
                    = \frac{1}{\eta^2}+\frac{1}{\eta}+e^{-\frac{1}{\eta}}.
\end{align}

According to \eqref{E_S}, \eqref{E_S^2} and \eqref{E_s_i}, the first two order moments of uplink service time $S_{\text{U}}$ are
\begin{align}
   \mathbb{E}(S_{\text{U}}) =&\mathbb{E}(S)\mathbb{E}(s_i)=\Big(1+\frac{\lambda \theta d^{\alpha}}{\rho}\Big) \Big(\frac{1}{\eta}+e^{-\frac{1}{\eta}}\Big),\\
   \mathbb{E}(S_{\text{U}}^2) =&\mathbb{E}(S)\mathbb{E}(s_i^2)+\mathbb{E}(S^2-S)\mathbb{E}^2(s_i)  \nonumber\\
                                =&\Big(1+\frac{\lambda \theta d^{\alpha}}{\rho}\Big)\Big(\frac{1}{\eta^2}+\frac{1}{\eta}+e^{-\frac{1}{\eta}}\Big)+  \nonumber \\
                                 & \left(\Big(\frac{\lambda \theta d^{\alpha}}{\rho}\Big)^2 +\frac{2 \lambda \theta d^{\alpha}}{\rho}\right) \Big(\frac{1}{\eta}+e^{-\frac{1}{\eta}}\Big)^2.
\end{align}

This completes the proof of \textit{Proposition} 2.
\end{proof}
\subsection{Proof of Theorem 2}
\begin{proof}
 We use a similar method to study the average downlink AoI as that in studying the uplink case.
    Based on \eqref{E_S} and \eqref{E_S^2}, we have the following result
\begin{align}
    \bar{\Delta}_\text{U}=& \frac{1}{\mathbb{E}(S_\text{U})} \Big(\mathbb{E}^2(S_\text{U}) + \frac{1}{2}\mathbb{E}(S_\text{U}^2) + \frac{1}{2}\mathbb{E}(S_\text{U})\Big) \nonumber \\
    =& \frac{3}{2} \Big(1+\frac{\lambda \theta d^{\alpha}}{\rho}\Big) \Big(\frac{1}{\eta}+e^{-\frac{1}{\eta}}\Big) +\frac{1}{2}+\frac{1}{2} \frac{1} {\eta+\eta^2 e^{-\frac{1}{\eta}}}  \nonumber \\
    & -\frac{1}{2} \frac{\rho}{\rho+\lambda \theta d^{\alpha}} \big(\frac{1}{\eta}+e^{-\frac{1}{\eta}}\big),
\end{align}
which proves \textit{Theorem} 2.

\end{proof}

\small{

\bibliographystyle{IEEEtran}

}

\end{document}